\documentclass[english,letterpaper,aps,prl,twocolumn,showpacs]{revtex4-1}
\usepackage{lmodern}

\usepackage[T1]{fontenc}
\usepackage[latin1]{inputenc}
\usepackage{graphicx}
\usepackage{amssymb}

\makeatletter

\usepackage{lmodern}
\usepackage{subfig}

\makeatletter

\usepackage{lmodern}

\makeatletter

\usepackage{lmodern}

\makeatletter

\usepackage{lmodern}

\makeatletter

\makeatletter

\usepackage{epsfig}

\usepackage{dcolumn}

\usepackage{bm}

\usepackage{bbm}

\usepackage{wasysym}

\newlength{\textwidthm}

\makeatother

\makeatother

\makeatother

\makeatother

\makeatother

\usepackage{babel}
\makeatother
\begin{document}

\title{Comment on the "Create Dirac Cones in Your Favorite Materials", by Chia-Hui Lin and Wei Ku (arXiv:1303.4822) }

\author{E. Kogan}
\email{Eugene.Kogan@biu.ac.il}

\affiliation{Department of Physics, Bar-Ilan University, Ramat-Gan 52900,
Israel}
\date{\today}

\begin{abstract}
Recently a paper by Lin and Ku \cite{lin} was posted, where  the authors propose a very interesting idea to engineer the Dirac points in   material which originally did  not have such points by inducing a CDW state through
introduction of impurities like vacancy, substitution, or
intercalation. In this comment we would like to explain the appearance of the Dirac points in such engineered  structures by symmetry arguments.
\end{abstract}


\maketitle

Recently a paper by Lin and Ku \cite{lin} was posted, where  the authors propose a very interesting idea to engineer the Dirac points in   material which originally did  not have such points. This is achieved by inducing a CDW state through
introduction of impurities like vacancy, substitution, or
intercalation.
In this comment we would like to explain the appearance of the Dirac points in such engineered  structures by symmetry arguments.

The authors of Ref. \cite{lin} actually propose to substitute the original atoms uniformly across the system in such a way, that a lattice with the basis  is obtained.
We will show below that Dirac points are feasible in any lattice with the basis which possess  the space-time inversion symmetry \cite{asano}.
The consideration throughout the comment is in the framework of a tight-binding model with the Hilbert space containing one orbital per atom.

To support our statement
let us recall first a well known case in which the basis contains two atoms.
Hence the Hamiltonian at the Bloch wave vector, ${\bf k} = (k_1,k_2)$, is
expressed by a $2\times $2 Hermite matrix, which can be expanded as
\begin{eqnarray}
\label{ham}
\hat{H}({\bf k})=E_0({\bf k})\hat{I}+{\bf R}({\bf k})\cdot{\bf \sigma},
\end{eqnarray}
where $E_0({\bf k})$ and ${\bf R}({\bf k})=(R_1({\bf k}),R_2({\bf k}),R_3({\bf k}))$ are are real
functions of ${\bf k}$, and $(\sigma_1, \sigma_2,\sigma_3)$ are the Pauli matrices. The contact of two bands, described by the Hamiltonian (\ref{ham}), takes place at ${\bf k}$, where
\begin{eqnarray}
\label{zero}
{\bf R}({\bf k})=0.
\end{eqnarray}
We have two unknowns, $k_1$ and $k_2$, and thus
Eq. (\ref{zero}) is overdetermined. To make the point contact
feasible, a constraint is required, which reduces the number of
conditions by one. Such a constraint may be supplied by a symmetry.

Let the Hamiltonian $\hat{H}({\bf k})$ possess the space-time inversion symmetry, that is
\begin{eqnarray}
\label{ham2}
\hat{H}({\bf k})=IT\hat{H}({\bf k})T^{-1}I^{-1}=\sigma_1\hat{H}^*({\bf k})\sigma_1,
\end{eqnarray}
where $I$ is the inversion operator and $T$ is the time-reversal operator. Substituting the Hamiltonian (\ref{ham}) into Eq. (\ref{ham2}) we obtain $R_3=0$, which makes the bands contact feasible.

The cases considered by Lin and Ku are more complicated. In particular, they
consider two examples of the square lattice; in one case, by substitution of  $1/3$ of the original atoms  uniformly across the system, a lattice with the basis containing 3 atoms is obtained, in another case,  by substitution of $1/4$ of the original atoms, a lattice with the basis containing 4 atoms is obtained (Figs. 3a and 4a of Ref. \cite{lin} respectively).

However, the simple mathematics presented above and treating the basis with two atoms, can be generalized for the case of any basis containing $n_S$ atoms and possessing the space-time inversion symmetry \cite{asano}. The point is that instead of
direct diagonalization of the Hamiltonian matrix $H({\bf k})$,  one could
renormalize it into smaller dimension following the formalism
of Brillouin and Wigner. If the $n_S$ dimensional
Hilbert space $S$ is divided  into $m$ dimensional subspace $S_A$ and $n_S - m$
dimensional subspace $S_B$, the $n_S\times n_S$ Hamiltonian matrix and its resolvent can be written as
\begin{eqnarray}
\label{ham5}
\hat{H}({\bf k})=\left(\begin{array}{ll} \hat{H}_{AA}({\bf k}) & \hat{H}_{AB}({\bf k}) \\  \hat{H}_{BA}({\bf k}) & \hat{H}_{BB}({\bf k}) \end{array} \right).
\end{eqnarray}
\begin{eqnarray}
\hat{G}({\bf k},\epsilon)=(\epsilon -\hat{H}({\bf k}))^{-1}=\left(\begin{array}{ll} \hat{G}_{AA}({\bf k}) & \hat{G}_{AB}({\bf k}) \\  \hat{G}_{BA}({\bf k}) & \hat{G}_{BB}({\bf k}) \end{array} \right).
\end{eqnarray}
The resolvent matrix in the subspace $S_A$ can be
written as
\begin{eqnarray}
\hat{G}_{AA}({\bf k},\epsilon)=(\epsilon -\hat{H}^{eff}({\bf k}))^{-1},
\end{eqnarray}
with  the energy dependent effective Hamiltonian presented by an $m\times m$ matrix
\begin{eqnarray}
\label{ham6}
&&\hat{H}^{eff}({\bf k},\epsilon)= \hat{H}_{AA}({\bf k})\nonumber\\
&&+ \hat{H}_{AB}({\bf k})\left[\epsilon-\hat{H}_{BB}({\bf k})\right]^{-1}\hat{H}_{BA}({\bf k}).
\end{eqnarray}
The full spectrum of the Hamiltonian (\ref{ham5}) can be found from  the energy dependent effective Hamiltonian (\ref{ham6}).

Looking at the Figs. 3a and 4a of Ref. \cite{lin} we realize that in both cases the basis possess  the inversion
symmetry, and the inversion is accompanied by the exchange of one pair of
atoms;  other atoms in the basis are not moved by the inversion (of course, two atoms separated by a lattice vector we treat as the same one). 
Taking the Hilbert space $S_A$ as that spanned by the $p_z$ orbitals of the pair of atoms exchanged by the inversion, we obtain Eq. (\ref{ham2}) in a modified form
\begin{eqnarray}
\label{ham22}
\hat{H}({\bf k})=IT\hat{H}({\bf k})T^{-1}I^{-1}=\tilde{\sigma}_1\hat{H}^*({\bf k})\tilde{\sigma}_1,
\end{eqnarray}
where
\begin{eqnarray}
\tilde{\sigma}_1=\left(\begin{array}{lll} 0 & 1 & 0 \\ 1 & 0 & 0 \\ 0 & 0 & 1 \end{array} \right)
\end{eqnarray}
for the  basis presented on Fig. 3a, and
\begin{eqnarray}
\tilde{\sigma}_1=\left(\begin{array}{llll} 0 & 1 & 0 & 0\\ 1 & 0 & 0 & 0 \\ 0 & 0 & 1  & 0\\ 0 & 0 & 0 & 1\end{array} \right).
\end{eqnarray}
for the  basis presented on Fig. 4a.
Thus $H_{AA}$ satisfies Eq. (\ref{ham2}), $H_{BB}$ satisfies equation
\begin{eqnarray}
\label{ham7}
IT\hat{H}_{BB}({\bf k})T^{-1}I^{-1}=\hat{H}_{BB}^*({\bf k}),
\end{eqnarray}
and $H_{AB}$ and  $H_{BA}$ satisfy equations
\begin{eqnarray}
\label{ham8}
IT\hat{H}_{AB}({\bf k})T^{-1}I^{-1}=\sigma_1\hat{H}_{AB}^*({\bf k})\nonumber\\
IT\hat{H}_{BA}({\bf k})T^{-1}I^{-1}=\hat{H}_{BA}^*({\bf k})\sigma_1.
\end{eqnarray}
Substituting Eqs. (\ref{ham2}), (\ref{ham7}) and (\ref{ham8}) into Eq. (\ref{ham6})
we recover the two-atoms basis case considered in the beginning of the comment, that is $\hat{H}^{eff}({\bf k},\epsilon)$ itself satisfies Eq. (\ref{ham2}). 
 
Thus we connect the existence of the Dirac points in the square lattice examples of Ref. \cite{lin} with the fact that Dirac points are feasible in any lattice with the basis which possess  the space-time inversion symmetry.

Discussion with Chia-Hui Lin is gratefully acknowledged.


\begin{thebibliography}{99}

\bibitem{lin} C.-H. Lin and W. Ku, arXiv:1303.4822.

\bibitem{asano} K.i Asano and C. Hotta, \prb {\bf 83}, 245125 (2011).

\end{thebibliography}
\end{document}